\documentclass[journal]{IEEEtran}
\usepackage{graphicx}
\usepackage{multirow}
\usepackage{bm}
\usepackage{subfigure}
\usepackage{amssymb}
\usepackage{algorithm}
\usepackage{algorithmicx}
\usepackage{algpseudocode}
\usepackage{amsthm}
\usepackage{setspace}
\usepackage{balance}

\newcommand{\y}{\boldsymbol{y}}

\newcommand{\h}{\boldsymbol{h}}

\newcommand{\A}{\mathcal{A}}
\newcommand{\I}{\boldsymbol{I}}

\newcommand{\MF}{\mathrm{MF}}
\newcommand{\BP}{\mathrm{BP}}

\newcommand{\CN}{\mathcal{CN}}

\newcommand{\bPhi}{\boldsymbol{\Phi}}

\newcommand{\balpha}{\boldsymbol{\alpha}}
\newcommand{\bomega}{\boldsymbol{\omega}}
\newcommand{\bgamma}{\boldsymbol{\gamma}}

\usepackage{cite}

\ifCLASSINFOpdf
\else
\fi

\usepackage[cmex10]{amsmath}
\begin{document}

\title{Low Complexity Sparse Bayesian Learning Using Combined BP and MF with a Stretched Factor Graph}

\author{Chuanzong~Zhang, Zhengdao~Yuan, Zhongyong Wang and Qinghua Guo%

\thanks{This work is supported by the National Natural Science
Foundation of China (NSFC 61172086, NSFC U1204607, NSFC 61201251). %It was also conducted in cooperation with the 4GMCT cooperative research project funded by Intel Mobile Communications Denmark ApS, Agilent Technologies, Aalborg University and the Danish National Advanced Technology Foundation.
}
\thanks{C. Zhang is with the School of Information Engineering, Zhengzhou University, Zhengzhou 450001, China, and the Department of Electronic Systems, Aalborg University, Aalborg 9220, Denmark (e-mail: ieczzhang@gmail.com).}
\thanks{Z. Yuan is with the Zhengzhou Institute of Information Science and Technology, Zhengzhou 450001, China
(e-mail: yuan\_zhengdao@foxmail.com).
}
\thanks{Z. Wang is with the School of Information Engineering, Zhengzhou University, Zhengzhou 450001, China (e-mail: iezywang@zzu.edu.cn).}
\thanks{Q. Guo is with the School of Electrical, Computer and Telecommunications Engineering, University of Wollongong, Wollongong, NSW 2522, Australia, and also with the School of Electrical, Electronic and Computer Engineering, University of Western Australia, Crawley, WA 6009, Australia (e-mail: qguo@uow.edu.au).}
}
\maketitle

\begin{abstract}
%\boldmath
%In this paper,  a sparse Bayesian learning (SBL) algorithm is proposed by iteratively implementing combined belief propagation (BP) and mean field (MF) message passing framework.  We design a stretched factor graph where \com{the time domain channel taps and frequency weights [the varibles in non-sparse and sparse domains]} are constrained by hard factors, so that BP, often showing advantage on performance, is used to deal with the hard constraint factors while MF is used for exponential factors. To further reduce the complexity, we approximately compute the BP messages by ignoring some minimal terms, yielding an approximate BP-MF SBL algorithm. Simulation results show that our proposed SBL algorithms have much lower complexity and slightly better mean-square-error (MSE) performance than MF SBL algorithm in vector-form, while outperform that in scalar-form with same order complexity.

This paper concerns message passing based approaches to sparse Bayesian learning (SBL) with a linear model corrupted by additive white Gaussian noise with unknown variance. With the conventional factor graph, mean field (MF) message passing based algorithms have been proposed in the literature. In this work, instead of using the conventional factor graph, we modify the factor graph by adding some extra hard constraints (the graph looks like being `stretched'), which enables the use of combined belief propagation (BP) and MF message passing.  We then propose a low complexity BP-MF SBL algorithm based on which an approximate BP-MF SBL algorithm is also developed to further reduce the complexity. Thanks to the use of BP, the BP-MF SBL algorithms show their merits compared with state-of-the-art MF  SBL algorithms: they deliver even better performance with much lower complexity compared with the vector-form MF SBL algorithm and they significantly outperform the scalar-form MF SBL algorithm with similar complexity. 
\end{abstract}


\IEEEpeerreviewmaketitle

\section{Introduction}\label{Sec:intro}
%\mdf{The introduction have been rewrited according to our discussion last time. Please...}
\IEEEPARstart{R}{ecently}, compressed sensing~\cite{Donoho2006,Candes2008} has received tremendous attention and it has found wide applications in a large variety of engineering areas, e.g. biomagnetic imaging, sparse channel estimation, bandlimited extrapolation and spectral estimation,  echo cancellation and image restoration~\cite{Wipf2004}.  In compressed sensing, a vector $\balpha\in\mathbb{C}^{L\times 1}$ which exhibits sparsity is estimated based on the measurement vector $\y\in \mathbb{C}^{N\times 1}$ with the following model
\begin{eqnarray}
\y=\bPhi\balpha+\bomega\label{eq:signalmodel}
\end{eqnarray}
where $\bPhi\in \mathbb{C}^{N\times L}$ is called dictionary matrix and $\bomega$ represents an additive white Gaussian noise (AWGN) vector with zero mean and  covariance matrix $\lambda^{-1}\I$.  In this work, we are particularly interested in the case that the variance of the AWGN (or the  precision parameter $\lambda$) is unknown. 

Besides convex~\cite{Chen1998} and greedy~\cite{Tropp2004} methods, sparse Bayesian  learning (SBL)~\cite{Tipping2001,Tipping2003,Shutin2011} is an alternative method of sparse signal estimation,  which aims at finding a sparse maximum a posteriori (MAP) estimate $\hat\balpha=\operatorname*{argmax}\limits_{\balpha} p(\balpha|\y)$ of the vector $\balpha$ by specifying a priori probability density function (pdf) $p(\balpha)$.  Instead of working directly with a prior  $p(\balpha)$, SBL typically employs a two-layer (2-L) hierarchical structure~\cite{Pedersen2015} that assumes a conditional prior pdf $p(\balpha|\bgamma)$ and a hyper-priori pdf $p(\bgamma)$, so that $p(\balpha)=\int_{\bgamma} p(\balpha|\bgamma) p(\bgamma)d\bgamma$ has a sparsity-inducing nature. Most recently, SBL has been efficiently implemented using belief propagation (BP) \cite{Tan2009,Baron2010} and approximate message passing \cite{Som2012,Al2014}. However, these methods assume that $\lambda$ is known, which may not be true in many applications. This work deals with message passing based approaches to SBL with unknown $\lambda$.  

Mean field (MF) based message passing~\cite{Xing2003,Bishop2003,Dauwels2007}, which is also often referred to as variational message passing (VMP), has been widely used for approximate Bayesian inference, especially for exponential distributions.
With 2-L or 3-L hierarchical priori structures, Pedersen et al. proposed an MF SBL algorithm  (with unknown $\lambda$)~\cite{Pedersen2012}, which was applied to sparse channel estimation in OFDM.  As the MF SBL algorithm deals with the sparse signal $\balpha$ in a vector-form, matrix inversion is involved in each iteration and its computational complexity is as high as $\mathcal{O}(L^3)$.  To address the issue of complexity,  a low complexity MF SBL algorithm [13] is then proposed, where the inverse of a large matrix is decomposed into a number of matrix inverses with smaller size.  Flexible trade-off between complexity and performance can be achieved by adjusting the size of smaller matrices, which means that the reduction of complexity comes at the cost of performance loss.  Apparently, the size of the smaller matrices can be set to be $1$, so that the matrix inverses are avoided and we call it scalar-form MF SBL algorithm.  Recently, the scalar-form MF SBL algorithm was used for channel gain and delay estimation in~\cite{Hansen2015}. We note that an efficient hyperprior $p(\balpha)$ with 2-L structure was proposed in~\cite{Tipping2001}, which performs better than the 2-L and 3-L structures in~\cite{Pedersen2012}. %\com{[cz: \cite{Hansen2015} uses the 2-L structure proposed by~\cite{Tipping2001} and we use the same 2-L structure. In our simulation, we find that the 2-L structure proposed~\cite{Tipping2001} performs better than both 2-L and 3-L structure in~\cite{Pedersen2012}, but \cite{Hansen2015} didn't mention this fact.]}

%Belief propagation (BP) is an alternative message passing technique for Bayesian inference. 
Different from MF which supposes all the beliefs of variable nodes are independent, BP considers the joint belief of variable nodes neighbouring a factor node and makes the most of their correlation. BP, which may achieve exact Bayesian inference, is efficient to deal with discrete probability models and linear Gaussian models. However, BP may have a high complexity, when especially dealing with models involving both discrete and continuous random variables. Recently, a unified message passing framework was proposed in~\cite{Riegler2013} where BP and MF are merged to keep the merits of BP and MF while avoid their drawbacks.

In this work, a low complexity BP-MF  SBL algorithm with a 2-L hierarchical prior is proposed. Instead of using the conventional factor graph shown in Fig.~\ref{fig:FG_a}, we modify the factor graph by adding a number of extra hard constraint factors as shown in Fig.~\ref{fig:FG_b}, i.e., the factor graph looks like being `stretched'.  The hard constraint factors seem redundant, which however facilitates the use of BP in the graph, leading to considerable performance improvement. As we assume that the noise variance $\lambda^{-1}$ is unknown, MF can be used to tackle the exponential factors, while BP is used to handle the hard constraint factors.  As we factorize the signal $\balpha$ in a scalar form, the developed BP-MF SBL algorithm avoids matrix inversion and has a low complexity. Inspired by the derivation of the generalized approximate message passing (GAMP)~\cite{Rangan2011}, we further simplify the BP message passing by ignoring some minimal terms and develop an approximate BP-MF SBL algorithm. Numerical examples show that the proposed BP-MF SBL algorithms provide even better mean-square-error (MSE) performance with much lower complexity compared with the vector-form MF SBL algorithm~\cite{Pedersen2012}, and achieve noticeable MSE performance gain with similar complexity compared with the scalar-form MF SBL algorithm~\cite{Pedersen2014,Hansen2015}.

\textit{Notation}- Boldface lowercase and uppercase letters denote vectors and matrices, respectively. %, while superscripts $(\cdot)^*$, $(\cdot){\tra}$ and $(\cdot){\herm}$ represent conjugation, transposition and Hermitian transposition, respectively.
 The expectation operator with respect to a pdf $g(x)$ is expressed by $\left\langle f(x) \right\rangle_{g(x)} = \int f(x) g(x) dx / \int g(x') dx' $, while $\textrm{var}[x]_{g(x)}= \left\langle \vert x\vert^2\right\rangle_{g(x)}-\vert\left\langle x\right\rangle_{g(x)}\vert^2  $ stands for the variance. The pdf of a complex Gaussian distribution with mean $\mu$ and variance $\nu$ is represented by $\CN(x;\mu,\nu)$. The relation $f(x)=cg(x)$ for some positive constant $c$ is written as $f(x)\propto g(x)$. %We use $\|\cdot\|$ to stand for Euclidian norm.

\section{Factor Graph Model}\label{Sec:SysMod}
\begin{figure*}[!t] \centering    
\subfigure[Conventional factor graph] { \label{fig:FG_a}     
\includegraphics[width=0.63\columnwidth]{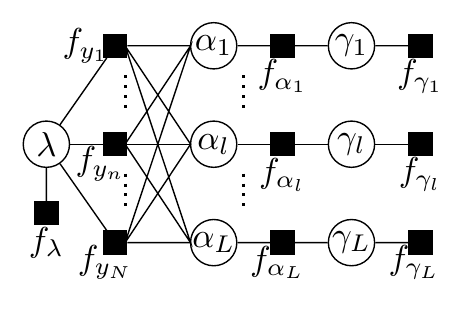}  
}     
\subfigure[Stretched factor graph] { \label{fig:FG_b}     
\includegraphics[width=0.8\columnwidth]{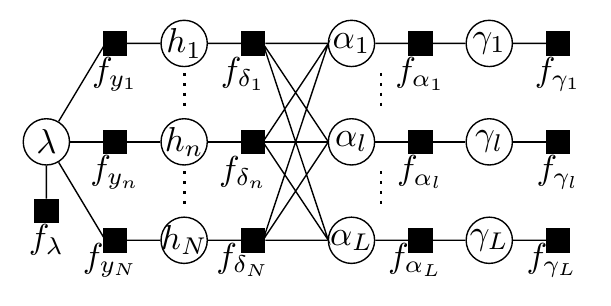}     
}     
\caption{Two factor graph representations for the probabilistic model \eqref{eq:factorization}.}     
\label{fig:Factor_Est}     
\end{figure*}

The joint a posteriori pdf of $\balpha,\bgamma$ and $\lambda$ in~\eqref{eq:signalmodel} with a 2-L hierarchical prior~\cite{Pedersen2015} can be factorized as
\begin{eqnarray}
p(\balpha,\bgamma,\lambda|\y)\propto{f_\lambda}(\lambda)\prod_n{f_{y_n}(\balpha,\lambda)} 
\prod_l{f_{\alpha_l}(\alpha_l,\gamma_l)f_{\gamma_l}(\gamma_l)},
\label{eq:factorization}
\end{eqnarray}
where  $f_{y_n}(\balpha,\lambda)\triangleq p(y_n|\balpha,\lambda)={\CN(y_n;\bPhi_n\balpha,\lambda^{-1})}$, with $\bPhi_n$ being the $n$-th row of matrix $\bPhi$, and $f_\lambda(\lambda)$ denotes the prior of noise precision parameter $\lambda$. %, and the factor $f_{\delta_n}(\balpha,\balpha)\triangleq p(h_n|\balpha)=\delta(h_n-\bPhi_n\balpha)$, where  $\bPhi_n$ is the $n$-th row of matrix $\bPhi$, stands for a deterministic (hard) constraint. %, $h_n=\bPhi_n\balpha$, between $h_n$ and $\balpha$.
The factor $f_{\alpha_l}(\alpha_l,\gamma_l)$ denotes the conditional pdf  $p(\alpha_l|\gamma_l)=\CN(\alpha_l;0,\gamma^{-1}_l)$, which is chosen  as a Gaussian prior of $\alpha_l$ and $f_{\gamma_l}(\gamma_l)$ represents a hyperprior $p(\gamma_l)=\mathrm{Ga}(\gamma_l;\epsilon,\eta)$\footnote{$\mathrm{Ga}(\cdot;a,b)$ denotes a Gamma pdf with shape parameter $a$ and rate parameter $b$. Note that, as in \cite{Tipping2001}, we use the Gama prior for the parameter of precision, rather than for variance~\cite{Pedersen2012}.} %Notice that the variance parameter $\gamma_l$ of $p(\alpha_l|\gamma_l) =\CN (\alpha_l;0,\gamma_l)$ is selected with a Gamma prior in its 2-L hierarchical priori structure~, while the precision parameter is selected with a Gamma prior in .
of the hyperparameter  $\gamma_l$. 
The factorization in (\ref{eq:factorization}) can be visually depicted on the factor graph~\cite{Kschischang2001} shown in Fig.~\ref{fig:FG_a}, which is similar to those in~\cite{Pedersen2014} and~\cite{Hansen2015}. We assume that $\lambda$ is unknown, and MF can be used to deal with factor nodes $\{f_{y_n},\forall n\in [1:N]\}$, which leads to the scalar-form MF SBL algorithm~\cite{Pedersen2014}. In~\cite{Pedersen2012}, the vector-form MF SBL algorithm is derived based on a conventional factor graph, where the vector $\balpha$ is treated as a single variable node.  %\com{[cz: uses a vector-form factor graph, where the vector $\balpha$ is a single variable node. Shall we mention it?]}

To facilitate the use of  both BP and MF, we modify the factor graph in Fig.~\ref{fig:FG_a} by adding hard constraint factors $\{f_{\delta_n}(h_n,\balpha)=\delta(h_n-\bPhi_n\balpha),\forall n\in [1:N]\}$ with a new variable vector $\h=\bPhi\balpha$. Therefore,  factor $f_{y_n}$ denotes the likelihood function $p(y_n|h_n,\lambda)=\CN (y_n;h_n,\lambda^{-1})$. The new factor graph,  shown in Fig.~\ref{fig:FG_b}, looks like a stretched version of the graph in Fig.~\ref{fig:FG_a}.  In the new graph, MF rules with fixed points equations can be used to compute the messages for the exponential factors, while BP rules, often yielding better performance, can be used to deal with the hard constraint factors. The message computations and scheduling are detailed in the following section.

\section{BP-MF Based SBL}\label{Sec:BPMF}
%\com{You can use the fonts of $\mathbb{A}\mathbf{B}\mathcal{C}\mathfrak{F}\mathit{G}\mathnormal{H}\mathrm{K}\mathsf{M}\mathtt{N}$}

In this section, with the combined BP-MF message update rule~\cite{Riegler2013}, we detail the message computations and scheduling on the factor graph shown in Fig.~\ref{fig:FG_b} to perform  sparse signal estimation. 
All the factors in Fig.~\ref{fig:FG_b} are represented by set $\A$, and it is divided into two disjoint subsets, a BP subset and an MF subset, which are denoted by $\A_{\BP}=\{f_{\delta_n},\forall n \}$ and $\A_{\MF}=\A\setminus\A_{\BP}$, respectively.

\subsection{ Message Computations}\label{sec:BP-MF}
%We will detail the messages computation from left to right (forward) at first, and then compute the message from right to left (backward). When the computation of forward messages needs the results of backward messages, we will exploit the messages  in previous iteration by default. 
The computations for messages passing from left to right (forward) and from right to left (backward)  are elaborated. The computations of some forward messages may need relevant backward messages, which we assume are produced  from the previous iteration.  
\subsubsection{Froward message computations}
Assuming that the belief $b(\lambda)$, later defined in (\ref{eq:b_lambda}),  of noise precision $\lambda$ is known, the message $m_{f_{y_n}\to{h_n}}(h_n)$ from observation factor $f_{y_n}\in\A_\MF$ to $h_n$ is calculated by the MF rule, as follows
\begin{eqnarray}
m_{f_{y_n}\to{h_n}}(h_n)&=&\exp\left\{\left<\log f_{y_n}(h_n,\lambda)\right>_{b(\lambda)}\right\}\nonumber\\
&\propto&\CN\left(h_n;y_n,\hat{\lambda}^{-1}\right),
\label{eq:fy2h}
\end{eqnarray}
where $\hat\lambda=\left\langle \lambda\right\rangle_{b(\lambda)}$.

The message $m_{f_{\delta_n}\to{\alpha_l}}(\alpha_l)$ from the hard factor $f_{\delta_n}\in\A_\BP$ to variable node $\alpha_l$ is computed by the BP rule with the messages  $n_{h_n\to{f_{\delta_n}}}(h_n)=m_{f_{y_n}\to{h_n}}(h_n)$ and $\{n_{\alpha_{l'}\to f_{\delta_n}}(\alpha_l') ,\forall l'\neq l\}$, later defined in~\eqref{eq:alpha2fh},  yielding
\begin{align}
m_{f_{\delta_n}\to{\alpha_l}}(\alpha_l)&=\left\langle {f_{\delta_n}}(h_n,\balpha)\right\rangle_{n_{h_n\to{f_{\delta_n}}}(h_n)\prod_{l'\neq l}n_{\alpha_{l'}\to f_{\delta_n}}(\alpha_{l'})} \nonumber\\
&\propto\CN\left(\alpha_l;\hat\alpha_{n\to l},{\nu_{\alpha_{n\to l}}}\right),
\label{eq:fh2alpha}
\end{align}
where
\begin{eqnarray}
\hat\alpha_{n\to l}&\triangleq&{\frac{y_n-{\hat{p}_n}+{\Phi_{nl}}{\hat\alpha_{l\to n}}}{\Phi_{nl}}}
\label{eq:fh2alpha_m}\\
\nu_{\alpha_{n\to l}}&\triangleq&{\frac{\hat{\lambda}^{-1}+\nu_{p_n}-|\Phi_{nl}|^2\nu_{\alpha_{l\to n}}} {|\Phi_{nl}|^2}}
\label{eq:fh2alpha_v}\\
{\hat{p}_n}&\triangleq &{\sum_l{\Phi_{nl}}{\hat\alpha_{l\to n}}}\label{eq:msg_p_m}\\
{\nu_{p_n}}&\triangleq&{\sum_l{|\Phi_{nl}|^2}{\nu_{\alpha_{l\to n}}}}\label{eq:msg_p_v}.
\end{eqnarray} %\com{[cz: Can you change the symbols such as $\nu^\alpha_{n\to l}$ and $\nu^p_{n}$ to $\nu_{\alpha_{n\to l}}$ and $\nu_{p_n}$, because we would append an iteration index in algorithm.]}
%For simplicity, we define two symbols,
%This operation should be repeated $NL$ times, get the message $m_{f_{\delta_n}\to{\alpha_l}}(\alpha_l)$, for $\forall l$ and $\forall n$.
%Since the messages $m_{f_{\delta_n}\to{\alpha_l}}(\alpha_l)$ fulfill Gaussian distribution, so we can use Gaussian multiplex formula to calculate message,

For convenience of description, the product of all the Gaussian messages $\{m_{f_{\delta_n}\to{\alpha_l}}(\alpha_l), \forall n\in[1:N] \}$ is denoted by  %$q_l(\alpha_l)$
\begin{eqnarray}
q_l(\alpha_l)&=&\prod_{n}{m_{f_{\delta_n}\to{\alpha_l}}}(\alpha_l)\nonumber\\
&\propto&{\CN\left(\alpha_l;\hat{q}_l,{\nu_{q_l}}\right)},
\label{eq:q_message}
\end{eqnarray}
where %\footnote{The product of two Gaussian functions is proportional to a Gaussian function, such as $\CN (x;\mu_1,\nu_1)\CN (x;\mu_2,\nu_2)\propto \CN (x;\mu,\nu)$ with $\nu =(1/\nu_1+1/\nu_2)^{-1}$ and $\mu =\nu (\mu_1/\nu_1+\mu_2/\nu_2)$.} %\com{[cz: Notice $q_l(\alpha_l)$ is not well-defined message, so we'd better not mention from A to B. Message $n_{\alpha_l \to f_{\alpha_l}}(\alpha_l)$ should be the belief, for $f_{\alpha_l}\in \A_\MF$. ]}
\begin{eqnarray}
{\nu_{q_l}} &\triangleq& \left(\sum_n\frac{1}{\nu_{\alpha_n\to l}}\right)^{-1}\label{eq:var_q}\\
%&\triangleq &{\left(\sum_n{\frac{|\Phi_{nl}|^2}{{{\hat\lambda^{-1}}}+{\nu_{p_n}}-{|\Phi_{nl}|^2}{\nu_{\alpha_{l\to n}}}}}\right)^{-1}} \\
{\hat{q_l}} &\triangleq& \nu_{q_l}\left(\sum_n\frac{\hat{\alpha}_{n\to l}}{\nu_{\alpha_{n\to l}}}\right)\label{eq:mean_q}.
%&\triangleq& {{\nu_{q_l}}\left(\sum_n{\frac{{\Phi_{nl}^*}({y_n}- \hat{p}_n+{{\Phi_{nl}}{\hat\alpha_{l\to n}}})}{{{\hat\lambda^{-1}}}+{\nu_{p_n}}-{|\Phi_{nl}|^2}{\nu_{\alpha_{l\to n}}}}}\right)}.
\end{eqnarray}
Given the  message $m_{f_{\alpha_l}\to{\alpha_l}}(\alpha_l)\propto  \CN \left(\alpha_l;0,{\hat\gamma_l}^{-1} \right)$, later defined in \eqref{eq:fGa2al}, the belief $b(\alpha_l)$ of variable $\alpha_l$ is obtained as
\begin{eqnarray}
b(\alpha_l)&\propto& q_l(\alpha_l)m_{f_{\alpha_l}\to{\alpha_l}}(\alpha_l)\nonumber\\
&\propto& \CN (\alpha_l;\hat{\alpha}_l,\nu_{\alpha_l}),
\label{eq:belief_alpha}
\end{eqnarray}
where
\begin{eqnarray}
\hat{\alpha}_l &\triangleq&\frac{\hat q_l}{1+\nu_{q_l} \hat\gamma_l} \label{eq:blf_alpha_m} \\
\nu_{\alpha_l} &\triangleq& \left({1/{\nu_{q_l}}+\hat\gamma_l}\right)^{-1}\label{eq:blf_alpha_v}.
%\mdf{\bar{\lambda}_l }&\triangleq& \mdf{\langle{\gamma_l^{-1}}\rangle_{b(\gamma_l)}}
\end{eqnarray}

%\com{[cz: we should give message $m_{f_{a_l}\to \gamma_l}$ in sequent.]}
Since the factor $f_{\alpha_l}$ is classified into the MF subset, the message $m_{f_{\alpha_l}\to\gamma_l}(\gamma_l)$ is calculated by using the MF rule,
\begin{eqnarray}
m_{f_{\alpha_l}\to\gamma_l}(\gamma_l)&=&\exp\left\lbrace\langle\log{f_{\alpha_l}}(\alpha_l,\gamma_l)\rangle_{b(\alpha_l)}\right\rbrace\nonumber\\
&\propto&\gamma_l\exp\left\{-\gamma_l(|\hat\alpha_l|^2+\nu_{\alpha_l})\right\},
\end{eqnarray}
so that  the belief $b(\gamma_l)$ of hyperparameter $\gamma_l$  reads 
\begin{eqnarray}
b(\gamma_l)&\propto& m_{f_{\alpha_l}\to\gamma_l}(\gamma_l) f_{\gamma_l}(\gamma_l)\nonumber\\
&\propto& \gamma_l^{\epsilon+1}\exp\left\{-\gamma_l(\eta+|\hat\alpha_l|^2+\nu_{\alpha_l})\right\}.\nonumber
\end{eqnarray}

%The updates of messages and beliefs in the prior part on left side of factor graph  are followed as \cite{Hansen2015}, and the computations of  the parameters, $\hat\gamma_l$, referred in \eqref{eq:fGa2al} read,
%\begin{align}
%\hat\gamma_l=\frac{\epsilon+1}{\eta+|\hat\alpha_l|^2+\nu_{\alpha_l}}
%\label{eq:gamma}
%\end{align}

\subsubsection{Backward Message}
We firstly compute the message  $m_{f_{\alpha_l}\to{\alpha_l}}(\alpha_l)$ from $f_{\alpha_l}$  to $\alpha_l$ by the MF rule, as follows
\begin{eqnarray}
m_{f_{\alpha_l}\to{\alpha_l}}(\alpha_l)&=&\exp \left\{\left\langle \log f_{\alpha_l}(\alpha_l,\gamma_l) \right\rangle_{b(\gamma_l)} \right\}\nonumber\\
&\propto & \CN \left(\alpha_l;0,{\hat\gamma_l}^{-1} \right), \label{eq:fGa2al}
\end{eqnarray}
where
\begin{align}
\hat\gamma_l=\left\langle \gamma_l\right\rangle_{b(\gamma_l)} =\frac{\epsilon+1}{\eta+|\hat\alpha_l|^2+\nu_{\alpha_l}}.
\label{eq:gamma}
\end{align}

%Then update the $\hat\gamma_l$ and $\hat\eta_l$ by
%$b(\gamma_l)={\gamma_l}^{\epsilon-2}{\exp(-\hat{\eta_l}\gamma_l)}$ and $b(\eta_l)={\eta_l^{\epsilon+a-1}\exp\{{-(\hat\gamma_l+b)\eta_l}\}}$ ((22) and (23) in \cite{Pedersen2012}).
%Thus can we use the updated $\hat\gamma_l$ to renew the belief $b(\alpha_l)$ by (\ref{eq:belief_alpha}).
Since factor node $f_{\delta_n}\in \A_{\BP}$, the message $n_{\alpha_l \to{f_{\delta_n}}}(\alpha_l)$ from variable node $\alpha_l$ to $f_{\delta_n}$  is updated by the BP rule,
\begin{eqnarray}
n_{\alpha_l \to{f_{\delta_n}}}(\alpha_l)&=&{\frac{b(\alpha_l)}{m_{f_{\delta_n}\to \alpha_l}(\alpha_l)}}\nonumber\\
&\propto& \CN({\alpha_l};{\hat\alpha_{l \to n}},{\nu_{\alpha{l\to n}}}),
\label{eq:alpha2fh}
\end{eqnarray}
where  %\com{[CZ: It seems we need the iteration index $t$. The expression of $\nu_{\alpha{l\to n}}$ includes itself in \eqref{eq:alpha2fh_v}. ]}
\begin{align}
{\nu_{\alpha_{l\to n}}}&\triangleq\left(\frac{1}{\nu_{\alpha_l}}-\frac{1}{\nu_{\alpha_{n\to l}}}\right)^{-1}\label{eq:alpha2fh_v}\\
%&\overset{\eqref{eq:fh2alpha_v}}{=}\left(\frac{1}{\nu_{\alpha_l}}-{\frac{|\Phi_{nl}|^2}{{{\hat\lambda^{-1}}}+{\nu_{p_n}}-|\Phi_{nl}|^2{\nu_{\alpha_{l \to n}}^{t-1}}}}\right)^{-1}\\
{\hat\alpha_{l \to n}}&\triangleq \nu_{\alpha_{l \to n}}\left(\frac{\hat\alpha_l}{\nu_{\alpha_l}} -\frac{ \hat{\alpha}_{n\to l}}{\nu_{\alpha_{n\to l}}}\right). \label{eq:alpha2fh_m}
%&\overset{\eqref{eq:fh2alpha_m}}{=}{\nu_{\alpha_{l\to n}}}\left(\frac{\hat\alpha_l}{\nu_{\alpha_l}}-{\frac{\Phi_{nl}^*(y_n-\hat{p}_n+\Phi_{nl}\hat\alpha^{t-1}_{l\to n})}{{{\hat\lambda^{-1}}}+{\nu_{p_n}}-|\Phi_{nl}|^2{\nu_{\alpha_{l \to n}}^{t-1}}}}\right).
\end{align}
%This process also needs to be repeated $PL$ times, get the message $n_{\alpha_l \to{f_{\delta_n}}}(\alpha_l)$ for $\forall l$ and $\forall n$.
Then the message $m_{f_{\delta_n}\to{h_n}}(h_n)$ can be  computed with  the BP rule for $f_{\delta_n}\in\A_\BP$, yielding
\begin{eqnarray}
m_{f_{\delta_n}\to{h_n}}(h_n)&=&\left\langle f_{\delta_n}(h_n,\balpha)\right\rangle _{\prod_l n_{\alpha_l \to f_{\delta_n}}(\alpha_l)}\nonumber\\
%&\propto{\CN\left({h_n};{\sum_l{\Phi_{nl}}{\hat\alpha_{l\to n}}},{\sum_l{|\Phi_{nl}|^2}{\nu_{\alpha_{l\to n}}}}\right)}\nonumber\\
&\triangleq&{\CN\left({h_n};{\hat{p}_n},{\nu_{p_n}}\right)}.
\label{eq:fh2h}
\end{eqnarray}
%where 
%\begin{eqnarray}
%\hat{p}_n \triangleq \sum_l{\Phi_{nl}}{\hat\alpha_{l\to n}};~~~~~\nu_{p_n}\triangleq \sum_l{|\Phi_{nl}|^2}{\nu_{\alpha_{l\to n}}}.
%\end{eqnarray}

We compute the belief $b(h_n)$ of variable $h_n$ by 
\begin{eqnarray}
b(h_n)&\propto&{m_{f_{\delta_n}\to{h_n}}}(h_n){n_{{h_n}\to{f_{\delta_n}}}}(h_n)\nonumber\\
&\propto& {\CN(h_n;{\hat{h}_n},{\nu_{h_n}})},\nonumber
\end{eqnarray}
where
\begin{eqnarray}
 \nu_{h_n} &\triangleq& \left({\hat{\lambda}+1/{\nu_{p_n}}}\right)^{-1} \label{eq:blf_h_v}\\
 \hat{h}_n &\triangleq& \nu_{h_n}\left({y_n\hat{\lambda}+\hat{p}_n/{\nu_{p_n}}}\right)\label{eq:blf_h_m}.
\end{eqnarray}
The message $m_{f_{y_n}\to{\lambda}}(\lambda)\propto\lambda{\exp\{-\langle{|y_n-h_n|^2}\rangle_{b(h_n)}\}}$ is calculated by the MF rule. With the conjugate prior pdf $f_{\lambda}(\lambda)\propto 1/\lambda$, the belief $b(\lambda)$  is updated by
\begin{eqnarray}
b(\lambda)&\propto& m_{f_{y_n}\to{\lambda}}(\lambda)f_\lambda(\lambda)\nonumber\\
&\propto&\lambda^{N-1}\exp\left\{-\lambda\sum_{n}\left\langle |y_n-h_n|^2\right\rangle _{b(h_n)}\right\}
\label{eq:b_lambda}
\end{eqnarray}
and  the parameter $\hat{\lambda}$ in \eqref{eq:fy2h} is computed as
\begin{eqnarray}
\hat\lambda=\langle{\lambda}\rangle_{b(\lambda)}
=\frac{N}{\sum_n{\langle{|y_n-{h_n}|^2}\rangle_{b(h_n)}}}
\label{eq:lambda}.
\end{eqnarray}

%\com{
%\subsubsection{\textit{Remark}} The main difference between the proposed BP-MF SBL and the scalar-form MF SBL algorithms lays the computation of message passed to variable node $\alpha_l$ from left hand side. We rewrite message $m_{f_{\delta_n}\to \alpha_l}(\alpha_l)$ in \eqref{eq:fh2alpha} on Fig.~\ref{fig:FG_b}, as
%\begin{align}
%& m_{f_{\delta_n}\to \alpha_l}(\alpha_l)\nonumber\\
%&=\int f_{\delta_n}(h_n,\balpha) n_{h_n\to f_{\delta_n}}(h_n) 
%\prod_{l'\neq l}n_{\alpha_{l'}\to f_{\delta_n}}(\alpha_{l'})d\alpha_{l'}dh_n\nonumber\\
%&=\frac{\int f_{\delta_n}(h_n,\balpha) n_{h_n\to f_{\delta_n}}(h_n) 
%\prod_{l'}n_{\alpha_{l'}\to f_{\delta_n}}(\alpha_{l'})\prod_{l'\neq l}d\alpha_{l'}dh_n}{n_{\alpha_l\to f_{\delta_n}}(\alpha_l)}\nonumber\\
%&=\frac{\int b(h_n,\balpha)\prod_{l'\neq l}d\alpha_{l'}dh_n}{n_{\alpha_l\to f_{\delta_n}}(\alpha_l)}
%\end{align}
%where $b(h_n,\balpha)$ denotes the joint belief of the variable nodes $\{\alpha_1,\cdots,\alpha_L,h_n\}$ neighbouring the factor nodes $f_{\delta_n}$. However, the vector-form MF SBL algorithm using the factor graph in Fig.~\ref{fig:FG_a} treats all the beliefs of variable nodes $\{\alpha_1,\cdots,\alpha_L\}$ as independent.
%}
\subsection{Message Scheduling  for BP-MF SBL Algorithm}
%\com{[descriptive sentences]}
The factors in Fig.~\ref{fig:FG_b} are very densely connected and thus there are a multitude of different
options for message scheduling. In this paper,  we simply choose a  schedule, where the messages are sequentially updated  in both forward and backward directions, while the messages in vertical direction are simultaneously computed for all $n\in[1:N]$ and $l\in[1:L]$. The BP-MF SBL algorithm with such scheduling is summarized in \textbf{Algorithm~\ref{alg:BPMF}}.
\begin{algorithm}
\caption{BP-MF SBL Algorithm}\label{alg:BPMF}
\begin{algorithmic}[1]
\State Initialize $\hat p_n$, $\nu_{p_n}$, $\hat\alpha_{l\to n}$, $\nu_{\alpha_{l\to n}}, \hat\gamma_l$, $\forall n$, $\forall l$ and $\hat\lambda$.
\For{$t=1\to \# $ of Iterations}
%\State $\forall n$: $m_{f_{y_n}\to{h_n}}(h_n)$ obtained by (\ref{eq:fy2h})
\State $\forall n,l$: update $\hat\alpha_{n\to l}$ and ${\nu_{\alpha_{n\to l}}}$ by (\ref{eq:fh2alpha_m}) and (\ref{eq:fh2alpha_v}).
\State $\forall l$: update  ${\nu_{q_l}}$ and $\hat{q}_l$  by (\ref{eq:var_q}) and (\ref{eq:mean_q}).
\State $\forall l$: update $\hat{\alpha}_l$ and $\nu_{\alpha_l}$  by (\ref{eq:blf_alpha_m}) and (\ref{eq:blf_alpha_v}).
\State $\forall l$: update $\hat\gamma_l$ by \eqref{eq:gamma}.%$ =\frac{\epsilon+1}{\eta+|\hat\alpha_l|^2+\nu_{\alpha_l}}$
\State $\forall l$: update $\hat{\alpha}_l$ and $\nu_{\alpha_l}$ again, by (\ref{eq:blf_alpha_m}) and (\ref{eq:blf_alpha_v}).
\State $\forall n,l$: update ${\nu_{\alpha_{l\to n}}}$ and  ${\hat\alpha_{l \to n}}$ by (\ref{eq:alpha2fh_v}) and (\ref{eq:alpha2fh_m}).
\State $\forall n$: update ${\hat{p}_n}$ and ${\nu_{p_n}}$  by (\ref{eq:msg_p_m}) and (\ref{eq:msg_p_v}).
\State $\forall n$: update ${\nu_{h_n}}$ and ${\hat{h}_n}$  by  (\ref{eq:blf_h_v}) and (\ref{eq:blf_h_m}).
\State update $\hat\lambda$ by \eqref{eq:lambda}, with $b(h_n)= \CN (h_n;\hat{h}_n,\nu_{h_n})$. %$ =\frac{N}{\sum_n{\langle{|y_n-{h_n}x_n|^2}\rangle_{b(h_n)}}}$
\EndFor\  $t$
\end{algorithmic}
\end{algorithm}
\section{Approximate BP-MF SBL}
It is observed that there are  $NL$ edges between variable nodes $\{\alpha_l,\forall l\}$ and factor nodes $\{f_{\delta_n},\forall n\}$, so we have to compute $2NL$ messages (see \textbf{Lines 3} and \textbf{8} in \textbf{Algorithm~\ref{alg:BPMF}}) for both forward and backward directions in each iteration. To simplify the BP-MF SBL, we  approximate the  means and variances of Gaussian messages in the BP part by eliminating some small terms, leading to the approximate BP-MF SBL algorithm.

%AMP algorithm primarily used in densely connected\footnote{Particular refer to edges between $\alpha_l$ and $f_{\delta_n}$ for $\forall l$ and $\forall n$ this paper} factor graph, can be interpreted as approximated BP rules, to reduce the computational complexity. The simplify of BP-MF algorithm using AMP rule reflected in the following three approximation expression:

\subsection{Approximation of Messages}
By substituting %\eqref{eq:mean_q} into 
\eqref{eq:blf_alpha_v} into \eqref{eq:alpha2fh_v}, 
\begin{eqnarray}
{\nu_{\alpha_{l\to n}}}=\left( {1}/{\nu_{q_l}}+\hat{\gamma}_l-{1}/{\nu_{\alpha_{n\to l}}}\right)^{-1}\approx{\nu_{\alpha_l}}\label{eq:l2n_v_LC}
\end{eqnarray}
can be obtained as ${1}/{\nu_{q_l}}\gg {1}/{\nu_{\alpha_{n\to l}}}$ from \eqref{eq:var_q} when the number $N$ is large enough. Similarly, substituting \eqref{eq:fh2alpha_m} and \eqref{eq:fh2alpha_v} into \eqref{eq:alpha2fh_m}, yields\footnote{To distinguish the parameters of messages in different iterations, we append a superscript $(t-1)$ to denote the index of the previous iteration. }
\begin{eqnarray}
%{\nu_{\alpha_{l\to n}}}&\thickapprox&{\nu_{\alpha_l}}\\
{\hat\alpha_{l\to n}} &=&{\nu_{\alpha_{l\to n}}}\left(\frac{\hat\alpha_l}{\nu_{\alpha_l}}-{\frac{\Phi_{nl}^*(y_n-\hat{p}_n+\Phi_{nl}\hat\alpha^{t-1}_{l\to n})}{{{\hat\lambda^{-1}}}+{\nu_{p_n}}-|\Phi_{nl}|^2{\nu_{\alpha_{l \to n}}^{t-1}}}}\right)\nonumber\\
&\approx&{\hat\alpha_l-{\nu_{\alpha_l}}\frac{y_n-\hat{p}_n}{\hat{\lambda}^{-1}+\nu_{p_n}}{\Phi_{nl}^{*}}}\nonumber\\
&=&{\hat\alpha_l}-{\nu_{\alpha_l}}{{s}_n}{\Phi_{nl}^{*}},
\label{eq:l2n_m_LC}
\end{eqnarray}
where 
\begin{equation}
{s}_n\triangleq{\frac{y_n-\hat{p}_n}{{\hat\lambda^{-1}}+\nu_{p_n}}}\label{eq:sfunction}.
\end{equation}
The above approximation is made by assuming that  the length $L$ of variable vector $\balpha$ is very large, so that $\hat{p}_n\gg\Phi_{nl}\hat\alpha_{l\to n}$ and $\nu_{p_n}\gg |\Phi_{nl}|^2\nu_{\alpha_{l\to n}}$ from \eqref{eq:msg_p_m} and \eqref{eq:msg_p_v}.

Substituting \eqref{eq:l2n_v_LC} and \eqref{eq:l2n_m_LC} into \eqref{eq:msg_p_v} and \eqref{eq:msg_p_m} respectively, we  obtain the approximate variance and mean 
\begin{eqnarray}
{\nu_{p_n}}&\approx &\sum_{l}{|\Phi_{nl}|^2}{\nu_{\alpha_l}}
\label{eq:msg_p_v_LC}\\
{\hat p_n}&\approx&{\sum_l{\Phi_{nl}}\left({\hat\alpha_l}-{\nu_{\alpha_l}}{s_n}{\Phi_{nl}^{*}}\right)}\nonumber\\
&\overset{\eqref{eq:msg_p_v_LC}}{\approx}&{\sum_l}{\Phi_{nl}}{\hat\alpha_l}-{s_n}{\nu_{p_n}}\label{eq:msg_p_m_LC}.
\end{eqnarray}
%If the length $L$ fo variable vector $\balpha$ is large enough, we consider $\nu_{p_n}\gg |\Phi_{nl}|^2\nu_{\alpha_{l\to n}}$ from \eqref{eq:msg_p_v},
We further substitute \eqref{eq:fh2alpha_v} and \eqref{eq:fh2alpha_m} into \eqref{eq:var_q} and \eqref{eq:mean_q}, and approximate them for a large $L$,  as follows,
\begin{eqnarray}
{\nu_{q_l}}&=&{\left(\sum_n{\frac{|\Phi_{nl}|^2}{{{\hat\lambda^{-1}}}+{\nu_{p_n}}-{|\Phi_{nl}|^2}{\nu_{\alpha_{l\to n}}}}}\right)^{-1}} \nonumber\\
&\approx&\left({\sum_n}{\frac{|\Phi_{nl}|^2}{{{\hat\lambda^{-1}}}+{\nu_{p_n}}}}\right)^{-1} \label{eq:msg_q_v_LC}\\
{\hat{q}_{l}}&=&{{\nu_{q_l}}\left(\sum_n{\frac{{\Phi_{nl}^*}({y_n}- \hat{p}_n+{{\Phi_{nl}}{\hat\alpha_{l\to n}}})}{{{\hat\lambda^{-1}}}+{\nu_{p_n}}-{|\Phi_{nl}|^2}{\nu_{\alpha_{l\to n}}}}}\right)}\nonumber\\
&\overset{\eqref{eq:sfunction}}{\approx}&{\nu_{q_l}}{\sum_n}\left( {\Phi_{nl}^{*}{s_n}}+{\frac{|\Phi_{nl}|^2}{{{\hat\lambda^{-1}}}+{\nu_{p_n}}}}\hat\alpha_{l\to n}\right)\label{eq:msg_q_m_LC1}\nonumber\\
&\overset{\eqref{eq:l2n_m_LC}\eqref{eq:msg_q_v_LC}}{\approx}&\hat\alpha_l+ {\nu_{q_l}}{\sum_n}\Phi_{nl}^{*}\left({s_n} - \frac{|\Phi_{nl}|^2}{{{\hat\lambda^{-1}}}+{\nu_{p_n}}}\nu_{\alpha_l}s_n\right)\nonumber\\
&\approx&{\hat\alpha_l}+ {\nu_{q_l}}{\sum_n}{\Phi_{nl}^{*}{s_n}}\label{eq:msg_q_m_LC}.
\end{eqnarray}
The approximation in \eqref{eq:msg_q_m_LC} is according to $\frac{|\Phi_{nl}|^2}{{{\hat\lambda^{-1}}}+{\nu_{p_n}}}\ll\nu^{-1}_{\alpha_l}$, since $\nu^{-1}_{\alpha_l}=\sum_n\frac{|\Phi_{nl}|^2}{{{\hat\lambda^{-1}}}+{\nu_{p_n}}}+\hat{\gamma}_l$ is obtained by inserting \eqref{eq:msg_q_v_LC} into \eqref{eq:blf_alpha_v}.

\subsection{Message Scheduling for Approximate BP-MF SBL Algorithm}\label{sec:Schedule}
%Based on the derivation presented in the last section, we scheduling the BP-MF and the A-BP-MF algorithms list in \textbf{Algorithm 1} and \textbf{Algorithm 2} respectively.
We choose the similar message scheduling to BP-MF SBL shown in \textbf{Algorithm~\ref{alg:BPMF}}, where the corresponding parameters are replaced by the above approximate computations. The parameters $\nu_{q_l}$ and $\hat q_l$ are updated by \eqref{eq:msg_q_v_LC} and \eqref{eq:msg_q_m_LC} instead of \eqref{eq:var_q} and \eqref{eq:mean_q}. The parameters $\nu_{p_n}$ and $\hat p_n$ are calculated by \eqref{eq:msg_p_v_LC} and \eqref{eq:msg_p_m_LC} rather than \eqref{eq:msg_p_v} and \eqref{eq:msg_p_m}. In addition, the computations of parameters $\hat{\alpha}_{n\to l}$, $\nu_{\alpha_{n\to l}}$, $\nu_{\alpha_{l\to n}}$ and $\hat{\alpha}_{l\to n}$  in \textbf{Lines 3} and \textbf{8} of \textbf{Algorithm~\ref{alg:BPMF}} are avoided, while a set of intermediate parameters $s_n$, $\forall n$, have to be inserted.   We summarize the approximate BP-MF SBL in~\textbf{Algorithm~\ref{alg:ABPMF}}. It is interesting that the message computations for the densely connected BP subgraph as shown in Fig.~\ref{fig:FG_b} coincide with the GAMP~\cite{Rangan2011} algorithm.

\begin{algorithm}
\caption{Approximate BP-MF SBL Algorithm}\label{alg:ABPMF}
\begin{algorithmic}[1]
\State Initialize %$\hat p_n$,
 $\nu_{p_n}$, $s_n$, $\forall n$, $\hat{\alpha}_l$, $\hat\gamma_l$,  $\forall l$ and $\hat{\lambda}$.
\For{$t=1\to \#$ of Iterations}
%\State $\forall n$: $m_{f_{y_n}\to{h_n}}(h_n)$ obtained by (\ref{eq:fy2h})
\State $\forall l$: update ${\nu_{q_l}}$ and $\hat{q}_l$  by (\ref{eq:msg_q_v_LC}) and (\ref{eq:msg_q_m_LC}).
\State $\forall l$: update $\hat{\alpha}_l$ and $\nu_{\alpha_l}$  by (\ref{eq:blf_alpha_m}) and (\ref{eq:blf_alpha_v}).
\State $\forall l$: update $\hat\gamma_l$ by \eqref{eq:gamma}.
\State $\forall l$: update $\hat{\alpha}_l$ and $\nu_{\alpha_l}$ again, by (\ref{eq:blf_alpha_m}) and (\ref{eq:blf_alpha_v}).
\State $\forall n$: update ${\nu_{p_n}}$ and  ${\hat{p}_n}$ by (\ref{eq:msg_p_m_LC}) and (\ref{eq:msg_p_v_LC}). 
\State $\forall n$: update $s_n$  by (\ref{eq:sfunction}).
\State $\forall n$: update ${\nu_{h_n}}$ and ${\hat{h}_n}$  by  (\ref{eq:blf_h_v}) and (\ref{eq:blf_h_m}).
\State update $\hat\lambda$ by \eqref{eq:lambda}, with $b(h_n)= \CN (h_n;\hat{h}_n,\nu_{h_n})$.
\EndFor\  $t$
\end{algorithmic}
\end{algorithm}

%\section{Compare BP-MF SBL Algorithm to Scalar-form MF SBL Algorithm}
%The two message passing algorithms are same on the update the messages about noise precision and the hyperparameter $\gamma$, so we only analyse the different part, dense-edges connection. On Fig.~\ref{fig:FG_a}, the message \begin{eqnarray}
%m_{f_{y_n}\to \alpha_l} = \exp\left\lbrace \left\langle \log f_y(\lambda,\balpha)\right\rangle _{b(\lambda)\prod_{l'\neq l}b(\alpha_{l'})}\right\rbrace 
%\end{eqnarray}
%where $b(\alpha_l)\approx p(\alpha_l|\y)$.
%
%We rewrite message $m_{f_{\delta_n}\to \alpha_l}(\alpha_l)$ in \eqref{eq:fh2alpha} on Fig.~\ref{fig:FG_b}, as
%\begin{align}
%& m_{f_{\delta_n}\to \alpha_l}(\alpha_l)\nonumber\\
%&=\int f_{\delta_n}(h_n,\balpha) n_{h_n\to f_{\delta_n}}(h_n) 
%\prod_{l'\neq l}n_{\alpha_{l'}\to f_{\delta_n}}(\alpha_{l'})d\alpha_{l'}dh_n\nonumber\\
%&=\frac{\int f_{\delta_n}(h_n,\balpha) n_{h_n\to f_{\delta_n}}(h_n) 
%\prod_{l'}n_{\alpha_{l'}\to f_{\delta_n}}(\alpha_{l'})\prod_{l'\neq l}d\alpha_{l'}dh_n}{n_{\alpha_l\to f_{\delta_n}}(\alpha_l)}\nonumber\\
%&=\frac{\int b(h_n,\balpha)\prod_{l'\neq l}d\alpha_{l'}dh_n}{n_{\alpha_l\to f_{\delta_n}}(\alpha_l)}
%\end{align}
%where $b(h_n,\balpha)\approx p(h_n,\balpha|y)$.
%
%Since MF algorithm assume all the belief $b(\alpha_l)$ are independent, while that BP rule considers the joint belief $b(h_n,\balpha)$ makes the most of their correlation.

\section{Numerical Simulation Results}\label{Sec:sim}
In this section, we assess the proposed SBL algorithms  by means of Monte Carlo simulations. %Consider an OFDM system with $M=1024$ subcarriers, out of which $N=100$ subcarriers are uniformly selected as pilot subcarriers. The pilot data $\x_p$ are set to ones and  the length of channel taps $\balpha$ is supposed to be $L=200$. So that the dictionary matrix $\bPhi$ is a sub-matrix selecting the rows of pilot subcarriers index and the first $L$ columns from an $M\times M$ discrete Fourier transform matrix $\F$~\footnote{The $(i,j)$-th entry of matrix $\F$ is $F_{i,j}=\exp\{-\jmath2\pi ij/M \}$.}.
%in signal model  denotes   We use a random $M\times N$ dictionary matrix $\bPhi$, whose entries are independent and identically distributed (IID) zero-mean complex Gaussian random variables with unit variance.
%We model the sparse channel, where the $K$ non-zero components uniformly dispersed in channel taps vector $\balpha$, and the non-zero taps are independent and identically distributed obeying a complex Gaussian distribution with zero-mean and unit variance.
%The simulated performance results of various algorithms are shown in Figs.~\ref{fig:MSEvsSNR},~\ref{fig:MSEvsIter}, and~\ref{fig:MSEvsSparse}, where BP-MF, A-BP-MF denotes,
Consider the sparse signal model~\eqref{eq:signalmodel} with a random $M \times N (M=100, N=200)$ dictionary matrix $\bPhi$, whose entries are independent and identically distributed (i.i.d.) zero-mean complex Gaussian random variables with unit variance. We assume that  the length-$N$ vector $\balpha$ has $K$ nonzero elements which are randomly dispersed in vector $\balpha$. In addition, the nonzero elements are i.i.d. and also drawn from a zero-mean complex Gaussian distribution with unit variance. All curves are produced based on  200 Monte-Carlo runs, and for each run with a new realization of the dictionary matrix $\bPhi$, the vector $\balpha$ and the AWGN vector $\bomega$ are generated. %The performance of the algorithm is evaluated based on the MSE of the estimator and the mean of number $\hat{K}$ of nonzero component in $\hat\balpha$.

We compare the MSE performance of our proposed  algorithms and the state-of-the-art algorithms. %, which we label as follows.
``BP-MF" and ``A-BP-MF" denotes our proposed BP-MF  and approximate BP-MF SBL  algorithms, i.e., \textbf{Algorithms~\ref{alg:BPMF}} and~\textbf{\ref{alg:ABPMF}}, respectively. ``MF-vector" and ``MF-scalar" stand for MF SBL algorithms in vector-form~\cite{Pedersen2012} and in scalar-form (sequentially estimating each element of the sparse signal $\balpha$)~\cite{Hansen2015}, respectively. For a fair companion, all the above algorithms use 2-L hierarchical structure with the hyperprior proposed in~\cite{Tipping2001}.  In addition, we also provide the performance of the vector-form MF algorithm using 3-L hierarchical prior in~\cite{Pedersen2012}, denoted by ``MF-vector-3L". %We encourage the interested readers to find more details in~\cite{Pedersen2012}.

\begin{figure}[!t]
\centering
\includegraphics[width=0.45\textwidth]{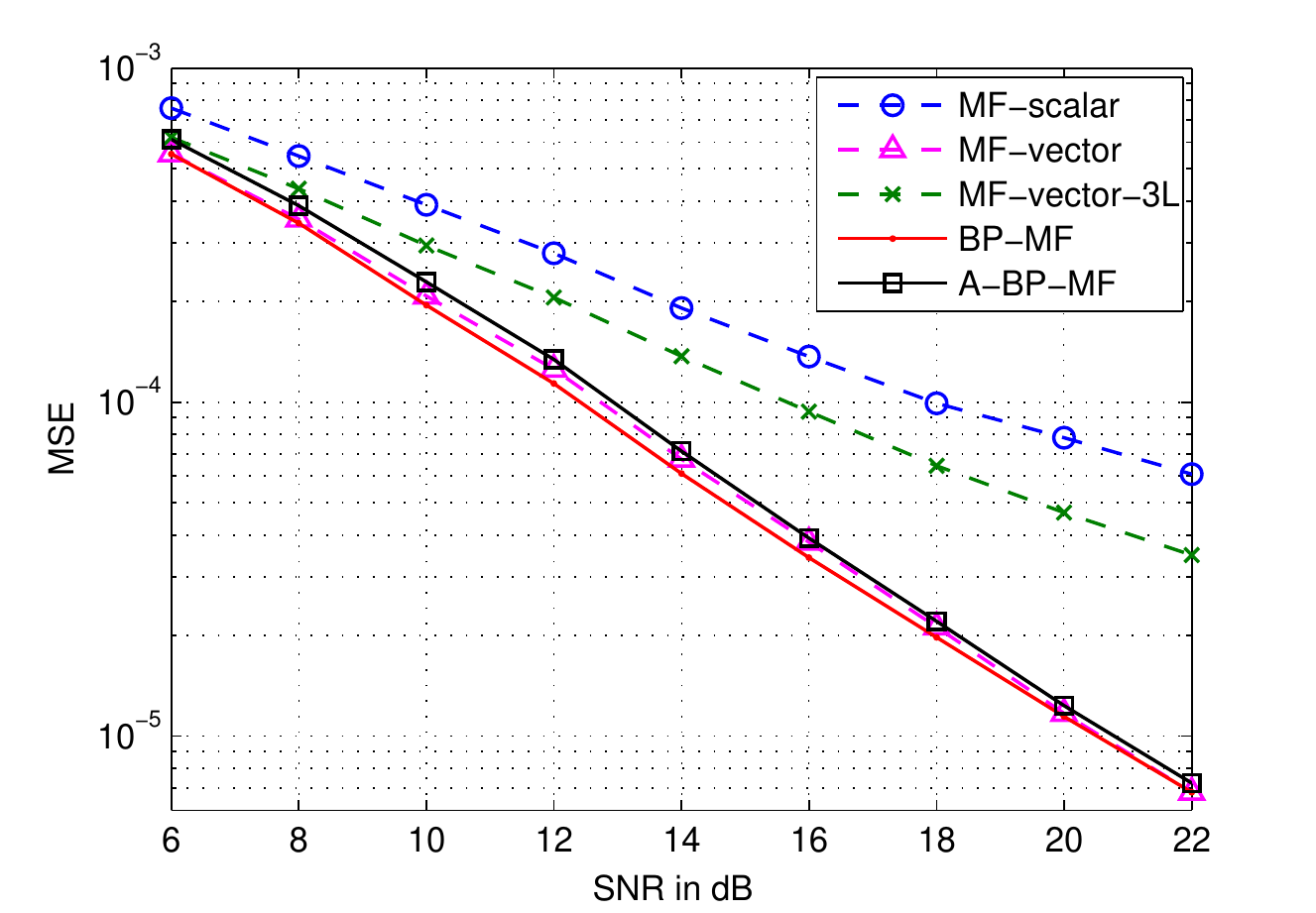}
\caption{MSE performance of different algorithms, where $K=26$.}\label{fig:MSEvsSNR}
\end{figure}

\begin{figure}[!t]
\centering
\includegraphics[width=0.45\textwidth]{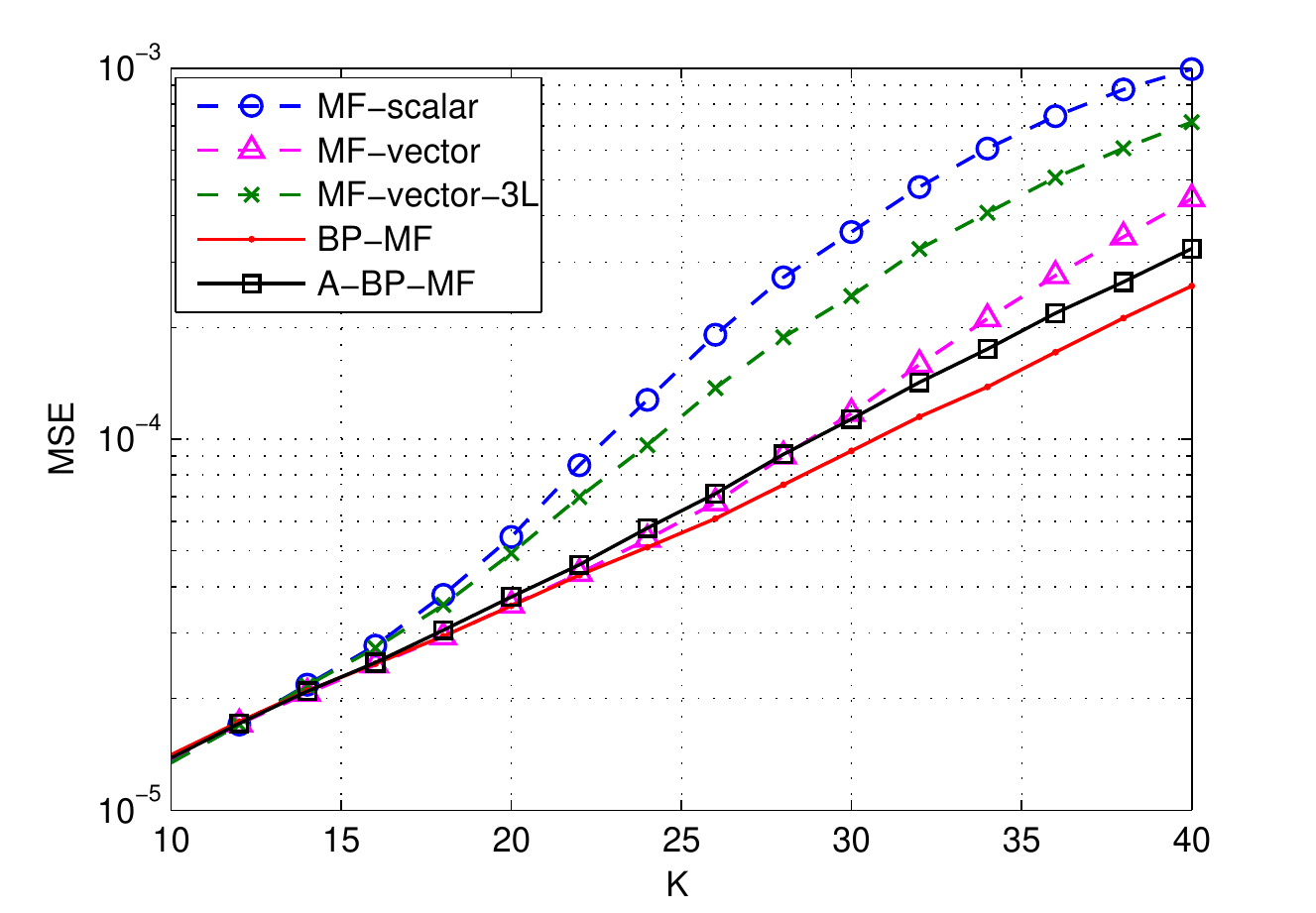}
\caption{MSE performance companions with number of nonzero components $K$, where SNR $= 14$dB. }\label{fig:MSEvsSparse}
\end{figure}

In Fig.~\ref{fig:MSEvsSNR}, the MSE performance of the algorithms is shown over a wide range of signal-to-noise ratios (SNRs), where all algorithms run 20 iterations and the number  of nonzero elements $K=26$.  We can observe that the proposed BP-MF and A-BP-MF algorithms  deliver slightly better MSE performance than MF-vector, and significantly outperform MF-scalar and MF-vector-3L.  
Fig.~\ref{fig:MSEvsSparse} depicts MSE performance with an SNR of 14dB  versus the number of non-zero elements  $K$. It shows that all the algorithms have similar performance when $K$ is small. However, with the increase of $K$, the MF SBL algorithms exhibit considerable performance loss compared to the proposed BP-MF and A-BP-MF SBL algorithms. It is also seen that BP-MF performs slightly better than A-BP-MF.

\begin{figure}[!t]
\centering
\includegraphics[width=0.45\textwidth]{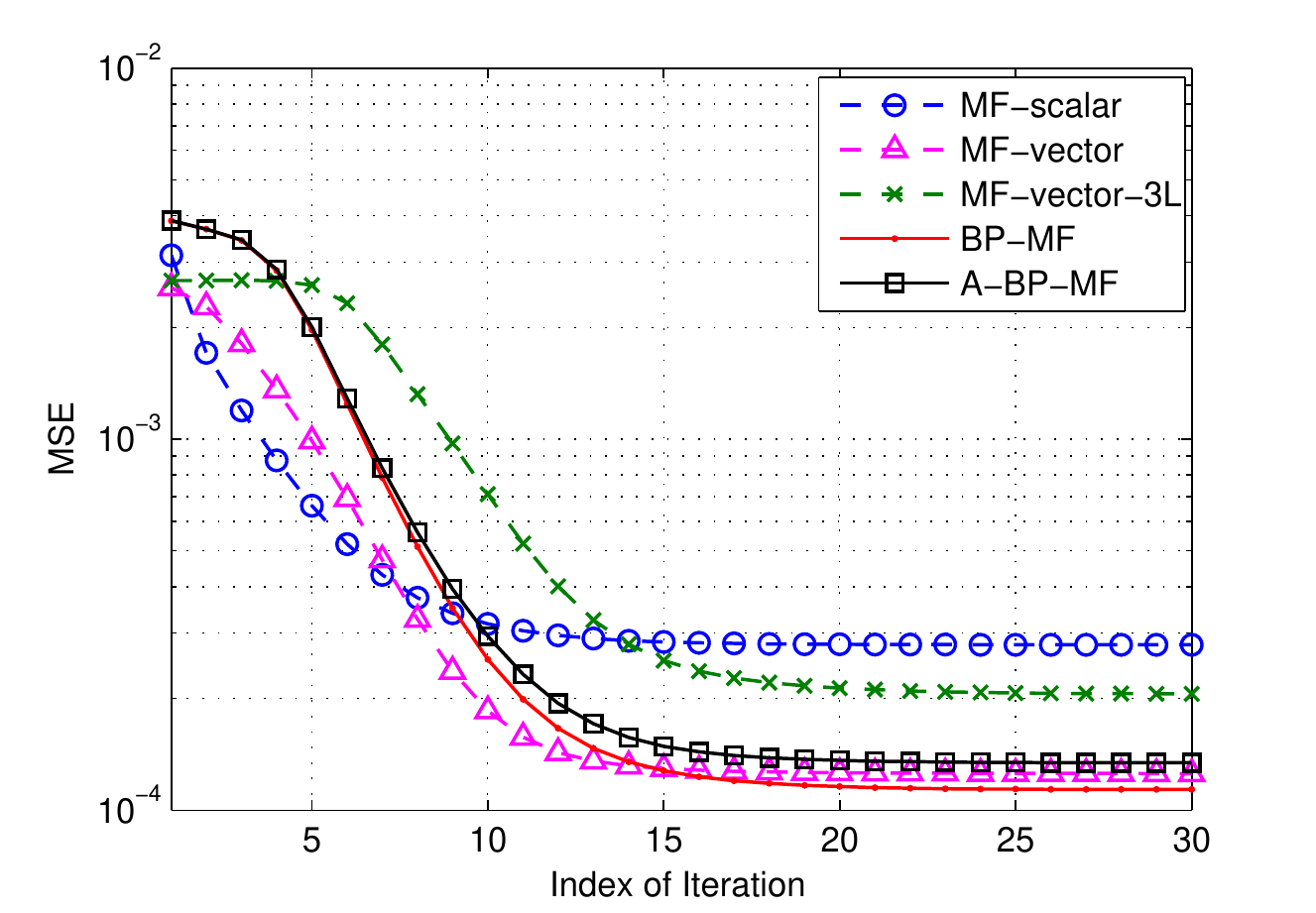}
\caption{MSE performance versus iteration index, where $K=26$ and SNR $=14$dB.}\label{fig:MSEvsIter}
\end{figure}

Fig.~\ref{fig:MSEvsIter} illustrates the convergence of the algorithms, where SNR $=14$dB and $K=26$. We can see that MF-scalar has the fastest convergence rate due to its sequential message updating schedule. Our proposed BP-MF algorithms converge slower but achieve better MSE performance compared to MF-salar and MF-vecotr-3L.  It can also be seen that our proposed algorithms have similar convergence rate and performance compared to MF-vector.

In addition, our simulation results in Figs.~\ref{fig:MSEvsSNR},~\ref{fig:MSEvsSparse} and \ref{fig:MSEvsIter} also show that the 2-L hierarchical priori structure proposed in~\cite{Tipping2001} outperforms 3-L hierarchical priori structure~\cite{Pedersen2012}. 
%All curves are based on total 200 times of Monte-Carlo runs, for each run new realization of the dictionary matrix $\bPhi$, the vector $\balpha$ and the AWGN vector $\omega$ are generated.
%The performance of the algorithm is evaluated based on the MSE of the estimator and the mean of number $\hat{K}$ of nonzero component in $\hat\balpha$.
%Results in terms of MSE versus signal-to-noise ratio (SNR) are given in Fig.\ref{fig:MSEvsSNR}, while the convergence of the MSE with the number of iteration is illustrated in Fig.\ref{fig:MSEvsIter}, the mean of $\hat{K}$ versus the SNR is plotted in Fig.4.
%Form Fig.2 we can find that the performance of all three algorithms is almost indistinguishable. The algorithm of BP-MF have faster convergence rate than others, while the AMP converge slow but have lowest computational complexity.

\subsection{Computational Complexity}
%This article is focusing on the investigation of reducing the complexity of ``MF-scalar" sparse Bayesian estimation algorithm in~\cite{Pedersen2012} involving the inversion of large matrices, 
As the message computations for updating $\lambda$ and $\gamma_l$ are the same for all the algorithms, we only  analyze the complexity of message computations related to $\h$ and $\balpha$. Due to the  matrix inversion involved, MF-vector has a complexity  of $\mathcal{O}(L^3)$ per iteration, while MF-scalar  $\mathcal{O}{(NL)}$. Since the proposed BP-MF and A-BP-MF algorithms using scalar-form factor graph shown in Fig.~\ref{fig:FG_b}, they have similar complexity to MF-scalar. In details, BP-MF needs to compute $\mathcal{O}{(NL)}$ messages %\mdf{[cz: There are $M\times N$ edges connecting $\{f_{\delta_n},\forall n\}$ and $\{\alpha_l,\forall l\}$ in Fig.~\ref{fig:FG_b}, so we have to compute $2MN$ messages in each iteration.]} 
and $\mathcal{O}{(NL)}$ memory cells to store the parameters (means and variances) of messages (see  \textbf{Lines 3} and \textbf{8} in \textbf{Algorithm~\ref{alg:BPMF}}), while MF-scalar and A-BP-MF only need to update and store $\mathcal{O}(N+L)$ messages. However, in updating the belief $b(\alpha_l), \forall l\in [1:L]$ MF-scalar with sequential message schedule may take longer running time than BP-MF algorithms. 

% The computation of each message $m_{f_{\delta_n}\to{h_n}}(h_n)$ and $n_{{\alpha_l}\to{f_{\alpha_l}}}(\alpha_l)$, need sum operation for $\forall n$ and $\forall l$ respectively. The BP-MF algorithm needs $16NL$ product operations every iteration. The A-BP-MF algorithm could omit the computation of the message passing in dense connection, but when it come to the calculation of message $m_{f_{\delta_n}\to{h_n}}(h_n)$ and $m_{{\alpha_l}\to{f_{\alpha_l}}}(\alpha_l)$, it still need $4NL$ product operations for each iteration. It means that the AMP method could reduce the complexity but not significantly.

\section{Conclusion}
In this paper, we  have investigated message passing based approaches to SBL. Two  low complexity BP-MF SBL algorithms have been proposed based on a stretched  factor graph which is obtained by adding extra hard constraint factors to the conventional factor graph. It has been shown that the BP-MF SBL algorithms outperform the state-of-the-art MF SBL algorithms in terms of computational complexity or performance.
%Using combined BP-MF message passing on factor graph, we present two low-complexity SBL algorithms performing in both non-sparse and sparse domain. A stretched factor graph in scalar-form is also proposed to implementing both MF and BP message update rules, which can take advantages of their virtues. The numerical evaluation results shows the advantages on MSE performance and computational complexity of BP-MF algorithms compared to those of MF algorithms. The performance gain may obtain from exploiting BP message update rule.

\ifCLASSOPTIONcaptionsoff
  \newpage
\fi

\balance

\end{document}